\documentclass[aps,twocolumn,showpacs,nofootinbib,superscriptaddress]{revtex4}
\usepackage{graphicx} 
\usepackage{amsmath} 
\usepackage{amsfonts} 
\usepackage{amssymb} 
\usepackage{hyperref} 
 
\newcommand{\ket}[1]{|{#1}\rangle} 
\newcommand{\bra}[1]{\langle{#1}|} 
 
\newcommand{\beq}{\begin{equation}} 
\newcommand{\eeq}{\end{equation}} 
\newcommand{\beqa}{\begin{eqnarray}} 
\newcommand{\eeqa}{\end{eqnarray}} 
 
\begin{document} 
 
\title{Decoherence of Bell states by local interactions with a dynamic spin environment} 
 
\author{Cecilia Cormick} 
\affiliation{Departamento de F\'\i sica, FCEyN, UBA, 
Ciudad Universitaria Pabell\'on 1, 1428 Buenos Aires, Argentina} 
 
\author{Juan Pablo Paz} 
\affiliation{Departamento de F\'\i sica, FCEyN, UBA, 
Ciudad Universitaria Pabell\'on 1, 1428 Buenos Aires, Argentina} 
 
\date{\today} 
 
\begin{abstract} 
We study the evolution of a system of two qubits, each of which interacts locally 
with a spin chain with nontrivial internal Hamiltonian. We present a new 
exact solution to this problem and analyze the dependence of decoherence on the distance between  
the interaction sites. In the strong coupling regime we find that decoherence increases with increasing distance. 
In the weak coupling regime the dependence of decoherence with distance is not generic (i.e., 
it varies according to the initial state). Decoherence becomes independent of distance when the latter is over a saturation length $l$. Numerical results for the Ising chain suggest that the saturation scale is related to the correlation length $\xi$. For strong coupling we display evidence of the existence of non--Markovian effects (such as 
environment--induced  interactions between the qubits). As a consequence the system 
can undergo a quasiperiodic sequence of ``sudden deaths and revivals'' of 
entanglement, with a time scale related to the distance between qubits.  
\end{abstract} 
 
\date{\today} 
\maketitle 
 
\section{Introduction} 
 
Understanding the process of decoherence \cite{Zeh-1973, PazZ00, Zurek03} is not only  
important from a fundamental point of view but also essential to design good error  
correction strategies to prevent the collapse of quantum computers \cite{NielsenC00}.   
When a composite system interacts with an environment, decoherence  
typically generates loss of entanglement between fragments.  
This process may depend on many details such as the nature of the system--environment 
interaction, the existence of non--trivial spatial correlations in the  
environment, the internal environmental dynamics, etc. In this  
paper we make a step towards understanding decoherence when different 
subsystems are locally coupled to a common environment with non--trivial internal 
dynamics and spatial correlations.  
 
In a previous publication \cite{CormickP-2007} we analyzed the decoherence induced on a single qubit by the  interaction with an environment formed by a spin chain with an XY Hamiltonian  (see \cite{CormickP-2007} for references on decoherence by spin environments). Here, we consider a system of two qubits, each of which interacts with a different site of an XY spin chain. We  
solve this model generalizing previous results and obtain exact expressions for the evolution of the reduced density matrix of the two qubits for a large family of initial states (that includes any of the four, maximally entangled, Bell states). Our work goes beyond previous studies extending the usual ``central qubit'' model, where each qubit is  
homogeneously coupled to all the sites of the chain \cite{jing-2006}.  
In fact, our results enable us to analyze the way in which decoherence and disentanglement  
depend on the distance between the interaction sites. In this context we will study two limiting cases: when both qubits interact with the same site in the chain, and when the interaction sites are widely separated.  More interestingly, we will investigate intermediate situations and examine the nature of the transition between the two limits.  We will show that the environment correlation length plays a crucial role in setting the typical length for which decoherence stops depending on distance. We will also analyze the way in which the evolution of the system depends on the strength of the coupling between the system and the environment.  
As we will see, weak and strong coupling regimes are drastically different.  
For strong coupling we will show that the chain provides a medium through  
which the qubits can coherently interact. In such a case, the  
entanglement between them may exhibit quasiperiodic events of ``sudden deaths'' and  
``sudden revivals'' \cite{yu-2004-93} with a time scale that depends on the  
distance between qubits.  
 
The paper is organized as follows: in Section \ref{sec:themodel} we  
introduce the model, defining the Hamiltonians for the  
system, the environment, and the coupling between them. We also present  
the main formulas we will use to determine the decay of quantum coherence. In  
Sections \ref{sec:weakcoupling} and \ref{sec:strongcoupling} we study the  
loss of entanglement between the qubits as a consequence of their  
interactions with the environment, in the cases of weak  
and strong coupling with the chain respectively. Finally, in Section \ref{sec:conclusions}  
we summarize our results.

\section{The model and its analytic solution} 
\label{sec:themodel} 
 
\begin{figure}[t] 
\begin{center} 
\includegraphics[width=0.3\textwidth]{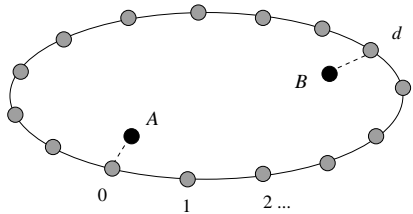} 
\caption{A system of two non-interacting qubits ($A$ and $B$), suffering decoherence  
from the local interaction with distant sites of an environment chain.} 
\label{fig:2qubitsystem} 
\end{center} 
\end{figure}

We study the decoherence induced on a system of two spin $1/2$ particles (the qubits)
by the coupling to an environment formed by a chain of $N$ spin $1/2$ particles  
(Fig. \ref{fig:2qubitsystem}). We neglect the self-Hamiltonian of the system.  
The total Hamiltonian is $H=H_C+H_{int}$, where $H_C$ is the Hamiltonian
of an XY spin chain: 
\beq 
H_C=-\sum_j\left\{ \frac{1+\gamma}{2} X_j X_{j+1} + \frac{1-\gamma}{2} Y_j Y_{j+1}  
+ \lambda Z_j\right\}. 
\eeq 
Here $X_j,Y_j,Z_j$ denote the three Pauli operators acting on the $j$-th site  
of the chain, and we assume periodic boundary conditions. The parameter $\gamma$  
determines the anisotropy in the $x-y$ plane and $\lambda$ gives a magnetic field in $z$  
direction ($\gamma=1$ corresponds to the Ising chain with transverse field).  
This model is critical for $\gamma=0$ with $|\lambda|<1$, and for $\lambda=\pm1$. 
 
Each of the two qubits in the system interacts  
locally with a certain spin of the chain: qubit $A$ interacts  
with spin $0$, and qubit $B$ with  
spin $d$. The interaction Hamiltonian is: 
\beq 
H_{int}=- g (\ket{1}\bra{1}_A Z_0 + \ket{1}\bra{1}_B Z_d) 
\eeq  
with $\ket{0}$ and $\ket{1}$ the two eigenstates of the $Z$ Pauli operator. 
This has a simple interpretation: the coupling to the qubits induces 
a change in the effective magnetic field at sites $0$ and 
$d$. Thus, if the system is in state $\ket{ab}$  
($a,b\in\{0,1\}$) the environment evolves with an effective  
Hamiltonian $H_{ab}$ given by: 
\beq 
H_{ab}=H_C - g (a Z_0 + b Z_d). 
\eeq 
 
We also assume that the initial state of the ``universe'' formed by 
system and environment is of the form: 
\beq  
\rho_{SE}(0)= \rho_0 \otimes \ket{E_0}\bra{E_0}  
\eeq  
where the initial state of the environment $\ket{E_0}$ is  
the ground state of the effective Hamiltonian $H_{00}$ \cite{Note0}.   

Our goal is to study the evolution of $\rho$, the reduced density matrix of  
the two-qubit system (obtained from the state of the universe by tracing  
out the environment). Because of the special form of the 
Hamiltonian, the temporal dependence of $\rho$ can be formally obtained 
as follows. In the basis of eigenstates of $Z_A$ and $Z_B$, 
$\rho$ can be written as: 
\beq 
\rho(t)=\sum_{abcd=0,1}\rho_{ab,cd}(t) \ket{ab}\bra{cd}. 
\eeq 
The evolution of the matrix elements of $\rho$ is given by: 
\beq 
\rho_{ab,cd}(t)=\rho_{ab,cd}(0) \bra{E_0}e^{iH_{cd}t} e^{-iH_{ab}t} 
\ket{E_0} 
\eeq 
(we take $\hbar=1$). As states in the basis commute with the total Hamiltonian, the diagonal  
terms $\rho_{ab,ab}$ remain constant. On the other hand, each  
off-diagonal term in the reduced density matrix is modified by a factor 
of absolute value between 0 and 1. This factor corresponds to the overlap between two different evolutions of the  
spin chain. Following \cite{LevsteinUP-1998, rossini-2007-75}, we denote the square modulus of this factor as the Loschmidt echo:  
\beq \label{eq:echodef} 
L_{ab,cd} (t) = |\bra{E_0}e^{iH_{cd}t} e^{-iH_{ab}t}\ket{E_0}|^2, 
\eeq 
which obviously satisfies $L_{ab,cd}=L_{cd,ab}$.  
 
Below,  
we will show how to compute the echoes $L_{00,ab}$ and 
$L_{01,10}$. These two echoes are enough to obtain the reduction  
of the off-diagonal elements in the reduced  
density matrix when the initial state is one of the  
four Bell states (or any state 
satisfying $\rho_{11,01}=\rho_{11,10}=0$).  
The  computation of each of these two echoes is slightly different.  
In fact, to obtain $L_{00,ab}$ we first note that one of the evolution  
operators in (\ref{eq:echodef}) acts trivially. Therefore,  
the echo is equal to the survival probability of the initial  
state after being evolved with the Hamiltonian $H_{ab}$, i.e. 
\beq 
L_{00,ab} (t) = \left|\bra{E_0} e^{-iH_{ab}t}\ket{E_0}\right|^2 
\eeq 
In turn, the echo $L_{01,10}$ can be calculated noticing  
that the corresponding effective Hamiltonians are  
related by a translation: 
\beq 
H_{01}=T^d H_{10} T^{-d}, 
\eeq 
where $T$ is the one--site translation operator in the chain. As the 
ground state of the Hamiltonian $H_{00}$ is an eigenstate 
of this operator, the echo can be written  
as: 
\beq \label{eq:echo0110} 
L_{01,10} (t) = \left|\bra{E_0} e^{iH_{10}t} T^d e^{-iH_{10}t}\ket{E_0}\right|^2 
\eeq 
 
To obtain the echoes we will make use of the fact  
that the full quantum evolution for each Hamiltonian $H_{ab}$ can be  
exactly solved. This can be done by mapping the Hamiltonians  
$H_{ab}$ of the chain onto a fermion system by means of  
the Jordan-Wigner transformation 
\cite{LiebSM-1961}: 
\beqa 
X_j&=&exp\left\{i\pi\sum_{k=1}^{j-1}c_k^\dagger c_k\right\} (c_j+c_j^\dagger)\\ 
Y_j&=&i~exp\left\{i\pi\sum_{k=1}^{j-1}c_k^\dagger c_k\right\} (c_j-c_j^\dagger)\\ 
Z_j&=&2c_j^\dagger c_j-1. \label{eq:transfZ} 
\eeqa 
Using this, up to a correction term associated to boundary effects, the Hamiltonians can be written 
as: 
\beqa 
H_{ab}&=&-\sum_j \big\{(c_j^\dagger c_{j+1} +c_{j+1}^\dagger c_j) + \gamma (c_j^\dagger c_{j+1}^\dagger 
+c_{j+1} c_j) +\nonumber\\ 
&&+ \lambda_j^{(ab)} (2c_j^\dagger c_j-1)\big\}  
\eeqa 
with $\lambda_j^{(ab)}=\lambda+ g (a\delta_{j,0}+b\delta_{j,d})$ (the extension 
to qubits interacting with more than one site is trivial).  
 
The Hamiltonians $H_{ab}$ depend quadratically on the annihilation and  
creation operators. Therefore they can be diagonalized by linear (Bogoliubov)  
transformations defining new creation and annihilation operators which we will  
denote as $\eta^{(ab)}, \eta^{\dagger(ab)}$. Furthermore, as all these  
transformations are linear, the operators corresponding to different  
values of the labels $(ab)$ can also be connected by Bogoliubov transformations. 
 
The echoes we want to compute can be written in terms of the matrices  
involved in these Bogoliubov transformations. For example, in Appendix A it 
is shown that: 
\beq \label{eq:mainformula} 
L_{00,ab} (t) = \left|\det({\rm g}+{\rm h} e^{i\Lambda t})\right|^2 
\eeq 
Here $\Lambda$ is a diagonal $N\times N$ matrix containing the energies of  
the normal modes of the Hamiltonian $H_{ab}$. ${\rm g}$ and ${\rm h}$  
correspond to the transformation connecting the particles that diagonalize  
the unperturbed Hamiltonian $H_{00}$ and the effective Hamiltonian 
$H_{ab}$:  
\beq \label{eq:lineartransf} 
\eta_j^{(00)}=\sum_k {\rm g}_{jk}\eta_k^{(ab)}+{\rm h}_{jk}\eta_k^{\dagger(ab)} 
\eeq 
This equation is useful since it expresses the echo as the  
determinant of an $N\times N$ matrix, which can be efficiently computed  
(the number of operations is polynomial in $N$). 
This formula is a new version of the one used  
in \cite{rossini-2007-75}, where the Loschmidt echo was written in  
terms of the two-point correlators of the environment chain (i.e.,  
as the determinant of a  $2N\times 2N$ matrix).  
 
The way to compute the echo $L_{01,10}$ is shown in Appendix B, using the fact  
that the translation operator $T$ is Gaussian 
in the fermion operators. The result is, once again, the determinant of  
an $N\times N$ matrix, though somewhat more complicated: 
\beq \label{eq:secondmainformula} 
L_{01,10} (t) = \left|\det({\rm g'} \tau^{-d} {\rm g'}^\dagger + {\rm h'} \tau^d {\rm h'}^\dagger)\right|  
\eeq 
Here $\rm{g'}$, $\rm{h'}$ are time-dependent complex matrices related to Bogoliubov transformations 
between different sets of particle operators, and $\tau$ is the diagonal matrix  
expression of the translation operator $T$. 
 
Decoherence for initial states of the form  
$|\varphi\rangle=\alpha\ket{00}+\beta\ket{11}$ or $|\varphi\rangle=\alpha\ket{01}+\beta\ket{10}$ is  
described by $L_{00,11}$ and $L_{01,10}$ respectively. In any case, the relevant echo $L(t)$ for the Bell-like state $|\varphi\rangle$ will 
determine not only the process of purity decay but also the way in which the 
two qubits become disentangled. In fact, if one considers an initial mixture of 
the form $\rho_0=p\ket{\varphi}\bra{\varphi}+(1-p)~\mathbb{I}/4$, the  
purity of the state is the following function of the echo $L(t)$:  
\beq 
Tr(\rho^2(t))=\frac{1-p^2}{2}+p^2[1-2|\alpha\beta|^2(1-L)]. 
\eeq 
(here, $L$ is either $L_{01,10}$ or $L_{00,11}$ depending on $|\varphi\rangle$).  
The entanglement between the two qubits 
can be measured by the negativity $\mathcal N$ obtained from  
the sum of the negative eigenvalues of the partial transpose of 
$\rho$ ($\mathcal N\neq 0$ is an indicator  of non-separability of 
the two-qubit state) \cite{VidalW-2002}. For the mixed initial state proposed:  
\beq 
\mathcal N(t)= \max\left\{0,p|\alpha\beta|\sqrt{L(t)} -\frac{1-p}{4}\right\}.  
\eeq 
For pure initial states ($p=1$), the state is entangled whenever $L\neq0$; in general,
the state becomes disentangled when the echo is  
below a threshold which depends on $p$. In this way the phenomenon 
of entanglement sudden death (ESD) may take place \cite{yu-2004-93}. 
 
We will analyze the dependence of the echo on the distance between the  
interaction sites. Two simple limits exist. First, if both qubits   
couple to the same site ($d=0$) then the echo $L_{01,10}$ becomes  
trivially 1 as the two effective Hamiltonians are identical. On the other  
hand, ${L}_{00,11}$ is simply the echo of a single qubit  
interacting with one site with twice the interaction strength (a case studied in \cite{rossini-2007-75}).  
The long distance limit is also easy to understand: if the chain is 
sufficiently large we expect its  effect to be equal to the one   
obtained when each qubit interacts with an independent environment.  
Then, both ${L}_{00,11}$ and ${L}_{01,10}$ approach ${L}_{0,1}^2$ (where ${L}_{0,1}$ is the single-qubit echo). 
This long distance regime exists only if the back-action of the  
qubits on the environment is  small. For strong back-action, one expects the  
environment to induce  effective interactions between 
the qubits. As we will see, the long distance limit will be identifiable easily in the  
weak coupling case, whereas the strong coupling regime will be plagued with  
effects associated to environment-induced interactions.

\section{Results: Weak coupling} 
\label{sec:weakcoupling} 
 
\begin{figure}[t] 
\begin{center} 
\includegraphics[width=0.45\textwidth]{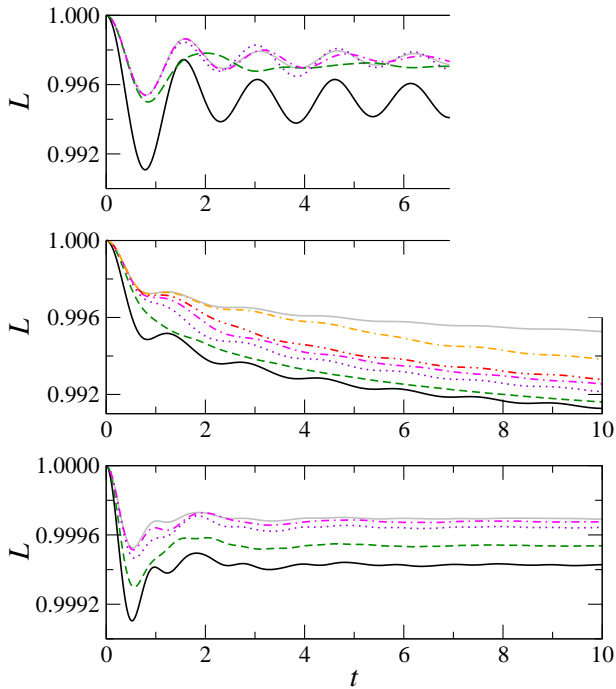} 
\caption{Echo $L_{00,11}$ as a function of time for an Ising chain ($\gamma=1$) of $N=100$  
sites weakly coupled to a two--qubit system ($g=0.1$). From top to bottom the 
transverse field is $\lambda=0.5$, $0.99$, and $1.5$. The distance between qubits  
is $d=0$ (full black), $1$ (dashed green), $2$ (dotted violet), and $3$ (dash-dotted magenta). In the almost  
critical case $\lambda=0.99$ we include also $d=4$ \mbox{($\cdot\cdot-$, red)},  
and $10$ \mbox{($--\cdot$, orange)}; these curves are not shown in the other plots because they are 
intertwined with the others. For comparison we display in full grey the limit of two independent 
environments.} 
\label{fig:peqpert_gamma1} 
\end{center} 
\end{figure}

In this Section we study the case when the coupling $g$ is small compared to the interaction strength 
between chain sites, showing results for the reference value $g=0.1$.   
\mbox{Fig. \ref{fig:peqpert_gamma1}} shows the echo $L_{00,11}$ as a function of time for 
the Ising chain ($\gamma= 1$) and for several values of the transverse field $\lambda$ and  
the distance $d$. 
For $\lambda<1$, the echo has small amplitude oscillations (with frequency of order $1$) about a  
value which is roughly independent of distance. For $\lambda>1$ decoherence is an order  
of magnitude smaller, oscillations decay rapidly and the echo approaches a constant that  
grows with distance. In both cases the dependence on distance rapidly saturates:  
the long distance regime is reached at $d\gtrsim 4$. To the contrary, near the critical point ($\lambda=1$)  
the echo decreases logarithmically with time (after a short transient), and saturation with distance is not attained, which is a signal of long range 
correlations in the environment.

\begin{figure}[t] 
\begin{center} 
\includegraphics[width=0.45\textwidth]{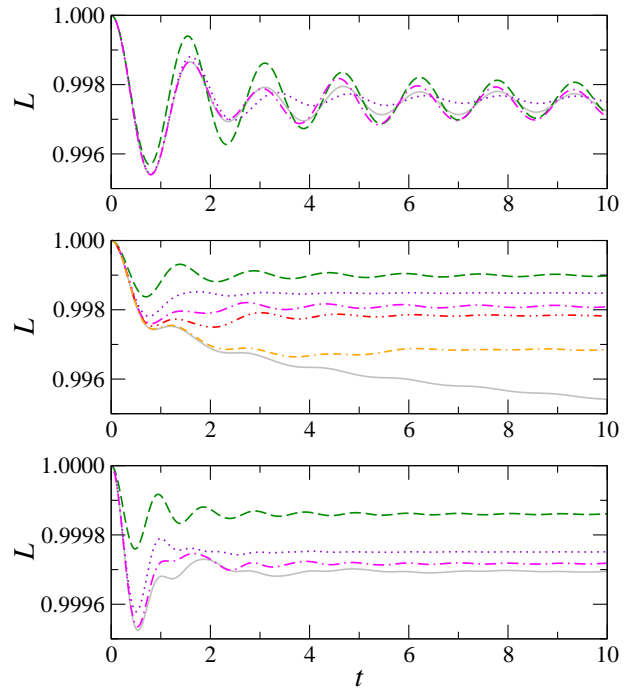} 
\caption{Echo $L_{01,10}$ as a function of time; all parameters are the same as in the previous Figure (for $d=0$ this echo is always equal to 1).} 
\label{fig:peqpert_gamma1_singlete} 
\end{center} 
\end{figure} 
 
The strongest decoherence is typically obtained at short distances. A simple argument shows that this is reasonable for weak coupling: 
at short times one expects an approximately Gaussian decay of the echo of the form $\exp(-\alpha^2t^2)$, with $\alpha\propto g$. For $d=0$ we have $\alpha^2\sim(2 g)^2$ while in the case of independent 
environments $\alpha^2\propto2 g^2$. As a consequence, faster decay of $L_{00,11}$ is expected for $d=0$ in the weak coupling regime. 
For the echo $L_{01,10}$ the dependence on distance is opposite to the previous case, as shown in 
Fig. \ref{fig:peqpert_gamma1_singlete}. This is expected since   
the Hamiltonians $H_{01}$ and $H_{10}$ become more different as $d$ grows.  
On the other hand, the saturation with distance approaching the limit of independent environments is similar to that of $L_{00,11}(g, t)$. 

\begin{figure}[t] 
\begin{center} 
\includegraphics[width=0.4\textwidth]{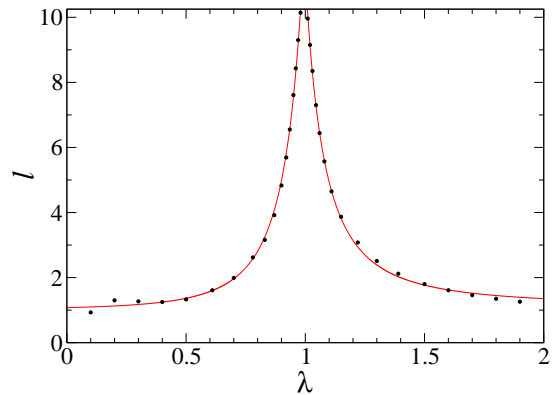} 
\caption{Saturation length as a function of the external field $\lambda$ (dots), for the Ising chain ($\gamma=1$). The continuous red line shows the fit as a function of the correlation length $\xi$ of the chain ($l\simeq1.1+0.21*(0.17+\xi^{-1})^{-2.2}$). These results correspond to a chain of 500 spins, and the echoes have been smoothed to avoid spurious effects due to oscillations. Besides, the fact that $g>0$ is accounted for by a shift in the correlation length, given by $\xi_{\rm eff}=\xi(\lambda+5.7~10^{-3})$.} 
\label{fig:saturation} 
\end{center} 
\end{figure}

We analyzed the saturation scale in the following way: we calculated the echo $L_{00,11}$ for $\lambda$ between $0.1$ and $2$, distances $d$ between $1$ and $15$, and times $t_j$ in the interval $[0,10]$. The squared norm of the vector obtained substracting the values of the echo at times $t_j$ for distance $d$ and for independent environments was taken as a measure of the difference between these echoes. A saturation length $l$ was obtained from the exponential fit of the decay of these norms as $d$ approaches the long distance limit. The behaviour of $l$ as a function of the external field $\lambda$ is shown in Fig. \ref{fig:saturation}. The saturation length was found to be related to the correlation length $\xi=|\ln(\lambda)|^{-1}$ \cite{Pfeuty-1970, BarouchMcCoy-1971}. Our numerical results indicate that $l\simeq1.1+0.21*(0.17+\xi^{-1})^{-2.2}$. Indeed, $l$ is clearly increased near the critical point, though it does not diverge like $\xi$. The analysis of the echo $L_{01,10}$ leads to analogous results.

The echoes obtained for the case \mbox{$\gamma=0.1$} are qualitatively similar to the ones shown in Figs. \ref{fig:peqpert_gamma1}, \ref{fig:peqpert_gamma1_singlete} for the Ising chain. There are nevertheless some noticeable differences. These concern mainly the dependence on distance, which is rather irregular for $\lambda\leq1$. Besides, the connection between the saturation distance $l$ and the correlation length $\xi$ is not evident, even though the saturation length clearly increases in the proximity of the critical point. There are also  differences in the shape of the decay close to criticality (almost linear for times up to $t\sim10$), and in the strength of the decoherence process (for $\lambda\leq1$ decoherence is stronger than in the case $\gamma=1$, while for $\lambda>1$ it is weaker). 

\begin{figure}[t] 
\begin{center} 
\includegraphics[width=0.45\textwidth]{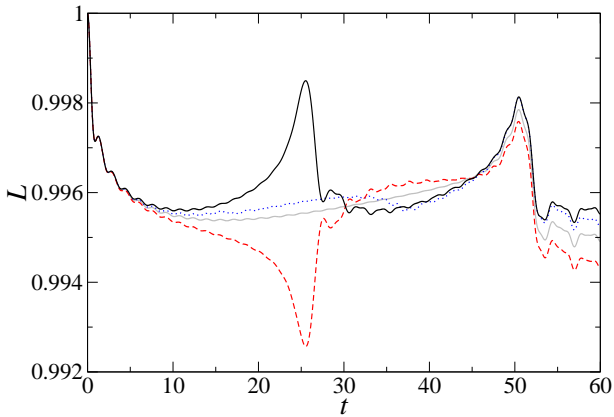} 
\caption{The echo for two qubits weakly interacting with a spin chain of  $N=100$ sites   
has revivals or sharp decays due to finite-size effects. For a distance $d=N/2$ we show $L_{00,11}$ (full black) and $L_{01,10}$ (dashed red). When $d$ decreases the first peak/decay tends to dissappear (the dotted blue line corresponds to $L_{00,11}$ for $d=30$). The result for two qubits coupled to  
independent environments is shown for comparison (full grey). The parameters are \mbox{$\lambda=0.99$}, $\gamma=1$, $g=0.1$.} 
\label{fig:1vs2chains} 
\end{center} 
\end{figure}

The previous results refer to times  
shorter than the ones where the finite size of the environment starts to play a role.  
For long times finite size effects become important, as shown by the coherence revivals and sharp decays of Fig. \ref{fig:1vs2chains}. These imply that the qubits are not independent due to 
environment--induced interactions. By decreasing $d$ below $N/4$ the  
first peak or decay tends to disappear, so that the qubits evolve almost independently  
even for long times (provided $d$ is over the saturation scale). This will not be the case in the strong coupling regime,  
which will be treated in the next section.  
 
 
\section{Results: Strong coupling} 
\label{sec:strongcoupling} 
 
For strong system-environment coupling,  
the results are quite different from the ones above. 
As observed for a single spin system in \cite{cucchietti-2006, CormickP-2007},  
the strong coupling regime is 
characterized by an echo with a fast oscillation and a slow envelope which for large  
enough $g$ is independent of $g$. 
Following the same steps as in \cite{CormickP-2007} we can explain the behaviour  
of the echo $L_{00,11}$ as follows:  
Consider the evolution with Hamiltonian $H_{11}=H_{00}-g (Z_0+Z_d)$. As all the frequencies associated with the interaction Hamiltonian are of order $g$ (much larger than typical frequencies of the chain), we may approximate: 
\beq 
e^{-iH_{11}t}\ket{E_0}\simeq e^{igt(Z_0+Z_d)} e^{-it H'}\ket{E_0}. 
\eeq 
Here $H'$ is the Hamiltonian $H_{00}$ reduced to a block diagonal form, with blocks  
associated with the different eigenvalues of the interaction term. The evolution  
operator thus factors in two parts: a fast periodic evolution with 
frequency of order $g$, and a slow evolution that determines the envelope of the echo and is governed by an effective chain Hamiltonian $H'$ (which does not depend on $g$). 
This same argument can be adapted to $L_{01,10}$. We note, however, that this behaviour is a consequence of the form of the interaction between system and environment and is not generic for strong couplings. For instance, if each qubit interacts with a region of the chain with a coupling that decreases with distance, the typical result is a sudden decay of coherence without oscillations.
 
\begin{figure}[t] 
\begin{center} 
\includegraphics[width=0.4\textwidth]{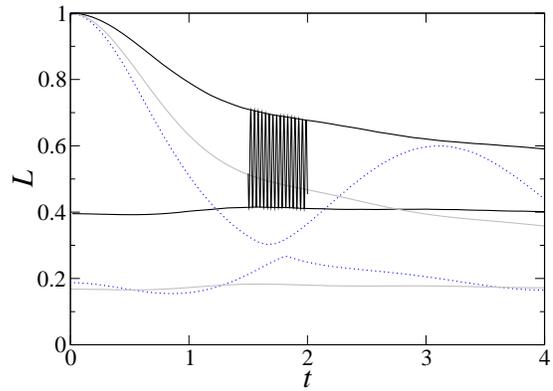} 
\caption{Envelope of the echo $L_{00,11}$ as a function of time for two qubits strongly interacting 
with two different sites of a chain with $N=100$, $\lambda=0.99$, $\gamma=1$, $ g=50$. The plots 
correspond to the cases of distance $d=0$ (full black), 2 (dotted blue), and long-distance limit 
(full grey). For the case $d=0$ a part of the fast oscillation is included. Here, in 
contrast to the weak coupling scenario, decoherence is stronger in the long distance limit, as the 
envelope for large $d$ corresponds to the square of the one for $d=0$.} 
\label{fig:revivals_gamma1} 
\end{center} 
\end{figure}

\begin{figure}[t] 
\begin{center} 
\includegraphics[width=0.4\textwidth]{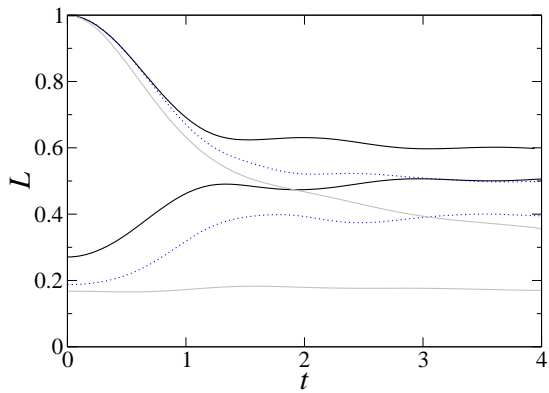} 
\caption{Envelope of the echo $L_{01,10}$ as a function of time for two qubits strongly interacting 
with two different sites of a chain with $N=100$, $\lambda=0.99$, $\gamma=1$, $ g=50$. The plots 
correspond to the cases of distance $d=1$ (full black), 2 (dotted blue), and long-distance limit 
(full grey). For the case $d=0$ there is no decoherence in this case, while for $d\geq1$ decoherence  
turns out to be stronger as distance is increased.} 
\label{fig:0110_largeg} 
\end{center} 
\end{figure}

For strong coupling we expect decoherence to increase with distance (until saturation) for both echoes.  
Indeed, for $d=0$ we have $L_{00,11}(g, t) = L_{0,1}(2 g, t)$, with $L_{0,1}$ 
the single qubit echo. This must 
be of the same order as  $L_{0,1}(g, t)$ because for large enough $g$ the envelope is  
independent of $g$. On the other hand, in the long 
distance limit $L_{00,11}(g, t) \simeq L_{0,1}^2(g, t)$, which is smaller than $L_{0,1}(g, t)$.  
The comparison between $d\gg1$ and $d=0$, illustrated in Fig. \ref{fig:revivals_gamma1} for $\gamma=1$, \mbox{$\lambda=0.99$}, thus leads to a result which is 
quite the opposite of the one obtained for weak coupling.  
The distance dependence of the echo $L_{01,10}$ follows a similar pattern, i.e., decoherence 
increases with distance, as shown in Fig. \ref{fig:0110_largeg}.  

Environment--induced interactions between qubits are manifest in  
Fig. \ref{fig:revivals_gamma1}. Indeed, for $d>0$ appart from the fast oscillation with a decaying  
envelope the echo $L_{00,11}$ exhibits a beating (or revival) at a time that depends 
on $d$. This effect is not important for weak coupling and can be interpreted as due 
to the interaction between the qubits through the modes of the perturbed chain.  
On the contrary, no revivals are seen for $L_{01,10}$. 

The difference can be  
understood by analyzing the origin of revivals in terms of the spectrum of  
the Hamiltonian and the Bogoliubov coefficients, contained in the matrices $\Lambda$, $\rm g$ and $\rm{h}$ in  (\ref{eq:mainformula}).  
Excitations of the unperturbed Hamiltonian have energies lying between 
$2|1-\lambda|$ and $2|1+\lambda|$.  
When a strong external field is applied in two sites, two eigenvalues of  
order $g$ appear; these excitations are associated to combinations of the original fermion operators in  
the two sites. The remaining excitations (with eigenvalues of the same order as before)
can be split in two groups, corresponding  
roughly to excitations between and outside the interaction sites (this labelling only makes sense because we consider distances $d\ll N$).  
It turns out that the most populated levels correspond to the lowest  
energy excitation (occupying the outside region), the excitations in  
the interaction sites, and those lying in the inside region \cite{Note1}. The 
high--energy excitations are associated to the rapid echo oscillation,  
while the lowest energy 
excitation has a time scale growing with $N$. The beating in the echo is given 
by the lowest-energy mode in the region between the qubits.  

The situation for  $L_{01,10}$ is different, as seen in Fig. \ref{fig:0110_largeg}, 
because for each effective Hamiltonian there is a single site perturbation of the chain. In this 
case the chain cannot be split in two regions and  
the populations of the low-energy levels decrease slowly with the energy, in  
such a way that there is not a single frequency associated to them. Then, no revivals 
occur up to times long enough for finite-size effects to appear. 
 
The revivals for the case $L_{00,11}$ appear for \mbox{$\lambda<1$} but not for 
\mbox{$\lambda>1$}. This is a  
consequence of the change in the properties of the chain. Indeed,  
for $\lambda<1$ the revival time is shorter as $\lambda$ is increased because the  
lowest energy associated to excitations between the qubits increases. For 
$\lambda>1$, the dissappearance of the revivals may be related to the fact that many  
low-energy excitations become populated, with frequencies comparable to the one associated to the beating. 

The first revival time, $t_r$, and the amplitude of the peak, $L_r$, can be analyzed as functions of the distance $d$. We consider first  
a critical case  where $H_{00}$ and $H_{11}$ belong to different phases ($\lambda=0.99$, $\gamma=1$, $N=100$). In each case, new peaks appear at multiples of $t_r$. For $d>1$, the revival time 
grows linearly with distance ($t_r \approx 2d$), while $L_r$ decays as a power law, $L_r\propto d^{-1/4}$. 
For distances $d \gtrsim N/5$ effects 
associated with the finite size of the chain appear altering the regular trend.  
Examining the echo for $\lambda$ far away from the phase transition 
we found a slower decay of the revival peak with $d$.  
Furthermore, in this case $t_r$ is not linear with $d$, but seems to 
grow exponentially. Thus, $t_r$ appears to be related to transport  
properties of the chain that are 
modified by varying the external field. This should serve as a warning not to 
picture revivals as the manifestation of a spin wave propagating with constant velocity 
along the chain (the energies of the lowest modes between the qubits do not generally 
scale as $1/d$). 
 
Changing the anisotropy parameter $\gamma$, the envelopes display different shapes,   
heights and revival times. The distance dependence of the revival time $t_r$ and amplitude  
$L_r$ was studied for $\lambda=0.99$ and  $\gamma$ between $0.1$ and $1$. We found  
that $t_r$ generally grows as a power law. Besides, the 
height of the revival peaks decreases with increasing $\gamma$, and the dependence  
of $L_r$ with $d$ maintains a power-law decay.
We note, however, that the peaks lose definition   
for large $d$ and small $\gamma$.

\section{Conclusions} 
\label{sec:conclusions} 
 
We studied the evolution of a system of two qubits interacting locally with  
an XY chain. Decoherence and disentanglement are determined by a Loschmidt echo,  
which we computed using previous results   
\cite{rossini-2007-75,cozzini-2006}. The formulas we obtained for the echoes enabled  us to study a family of initial states including the four Bell states.  
We focused our analysis on the dependence of the echo with the distance between the interaction sites. 
The cases we studied show a rich variety of results. For instance, for strong coupling decoherence increases with increasing distance for all Bell states. On the contrary, for weak coupling decoherence may increase or decrease with distance according to  the initial state chosen.  

In the regime of weak system--environment coupling
the dependence of the echo with distance saturates very fast except for the  critical case. The saturation length $l$ characterizing the approach to the long distance limit seems to depend on the correlation length of the chain $\xi$. This result is interesting as it relates an equilibrium quantity (the correlation length $\xi$) with a dynamical quantity $l$. However, this relation was only found for the Ising case, and will be further analyzed elsewhere. 

The strong coupling regime is  characterized by fast oscillations with a slowly decaying envelope. For some initial states, this envelope shows a beating which can be interpreted as an  
interaction of the qubits through the chain. Its time scale is determined by the  
modes of the chain in the region between the qubits, which depend on the distance between them and on the values of the parameters in the chain Hamiltonian (indeed, the beatings were found only for $\lambda<1$). We note that in this regime the fast oscillation of the echo  
continually provokes decays and revivals of the entanglement between qubits. This  
is just a dynamical transfer of this entanglement back and forth from the system to  
its immediate vicinity with a frequency given by the coupling constant $g$. The  
true loss of entanglement is produced by the decay of the  
oscillation. If the initial state is mixed, this decay can lead to true  
sudden death of entanglement. When the distance between qubits is  
short, the sudden death may be followed by a sudden revival due to the beating in the echo. For  
very long distances, entanglement loss becomes irreversible, unless  
we consider times which are long enough for the finite size of the  
environment to become manifest. 
 
Finally, it is worth pointing out that there are formal analogies between the model we solved and other decoherence models. In fact, decoherence for two qubits at distance $d$ in an initial state of the form $\alpha\ket{00}+\beta\ket{11}$ is identical to that of a single qubit interacting non-locally with sites $0$ and $d$ (which was partially studied in \cite{rossini-2007-75}). On the other hand, for the case of states of the form $\alpha\ket{01}+\beta\ket{10}$, the relevant echo $L_{01,10}$ can be mapped onto the one of another equivalent problem: a spinless particle occupying discrete positions along the chain (with position eigenstates $\ket{x_j}$, $j=0,\ldots,N-1$). If this particle modifies the effective magnetic field for the spin in the site it occupies, cat states of the form $(\ket{x_j}+\ket{x_{j+d}})/\sqrt{2}$ will decohere as a consequence of the interaction. The resulting reduction of the off-diagonal terms will be precisely given by expression (\ref{eq:secondmainformula}). 

\textit{Note added:} After the submission of this paper, we became aware of the work in \cite{rossini-2008-77}, which shows how to calculate efficiently the remaining elements of the density matrix ($\rho_{01,11}$).
 
\section{Appendix A} 
 
In this section we show how the formula (\ref{eq:mainformula}) for the echo can be obtained. Taking into account the relation (\ref{eq:lineartransf}) between the operators that diagonalize the different effective Hamiltonians, 
\beq 
\eta_j^{(0)}=\sum_k{\rm g}_{jk}\eta_k^{(1)}+{\rm h}_{jk}\eta_k^{\dagger(1)} \nonumber 
\eeq 
the two vacuum states $\ket{E_0}$ (for the unperturbed Hamiltonian) and $\ket{E_0'}$ (for the perturbed one) can be  
connected by \cite{chung-2001-64}: 
\beq \label{eq:vacuumrelation} 
\ket{E_0} \propto \exp\left\{\frac{1}{2} \vec \eta^{\dagger(1)} G \vec \eta^{\dagger(1)}\right\} \ket{E_0'}  
\eeq 
with $G=-{\rm g}^{-1}{\rm h}$ (for the sake of simplicity, we shall first assume that $\rm{g}$ is invertible, and sketch the most general  
case afterwards). The echo can then be calculated from: 
\beqa  
L (t) &=& |\bra{E_0} e^{-iH_1t} \ket{E_0}|^2 \nonumber\\ &\propto& |\bra{E_0'} e^{-\frac{1}{2} \vec \eta G^* \vec \eta} e^{-it\vec\eta^ 
\dagger\Lambda\vec\eta} e^{\frac{1}{2} \vec \eta^\dagger G \vec \eta^\dagger} \ket{E_0'}|^2 
\eeqa 
where in the last expression $\Lambda$ is a diagonal matrix containing the energies corresponding to the different particles, and all  
super-indices are omitted as all operators and matrices refer to the perturbed Hamiltonian $H_1$. By introducing two identities in terms  
of fermionic coherent states between the exponentials and integrating two times, \cite{NegeleO-1987} we obtain: 
\beqa  
L (t) &\propto& \Bigg|\int d\alpha_1\ldots d\alpha_N d\beta_1\ldots d\beta_N \nonumber\\ 
&& \exp\left\{\frac{1}{2} \vec \alpha G^* \vec \alpha+\vec\beta e^{it\Lambda}\vec\alpha-\frac{1}{2} \vec \beta G \vec \beta\right\} 
\Bigg|^2 
\eeqa 
where $\vec\alpha, \vec\beta$ are Grassman $N$-tuples. This is a Gaussian integral, that can be solved to: 
\beq \label{eq:Gaussianinttodet}  
L (t) \propto \left|\det 
\begin{pmatrix} 
G^* &-e^{it\Lambda}\\ 
e^{it\Lambda} & -G 
\end{pmatrix} 
\right| 
\eeq 
 
Using properties of the determinant and the fact that $\Lambda$ is diagonal, and imposing  
that $L(0)=1$ the echo can be rewritten as:  
\beq \label{eq:mainformula_previousstep} 
L_{01,10} (t) = \left|\det({\rm g} e^{it\Lambda} {\rm g}^\dagger + {\rm h} e^{-it\Lambda} {\rm h}^\dagger)\right| 
\eeq 
The formula can be simplified further: since $\rm{g}$, $\rm{h}$ are real we have that 
$\rm{g}\pm \rm{h}$ is orthogonal, and using this we obtain the final expression (\ref{eq:mainformula}): 
\beq  
L (t) = \left|\det({\rm g}+{\rm h} e^{i\Lambda t})\right|^2 \nonumber 
\eeq 
 
In case $\rm{g}$ is not invertible, the relation (\ref{eq:vacuumrelation}) between the vacuum states can be generalized by using  
intermediate sets of operators \cite{cozzini-2006}. By the singular value decomposition, ${\rm g}=UDV$ with $D$ diagonal, $U,V$  
orthogonal. We define new fermionic operators $\vec\xi_j^{(0)}=U^t\vec\eta^{(0)}$, $\vec\xi^{(1)}=V\vec\eta^{(1)}$. The  
linear transformation between the $\xi^{(i)}$, $\xi^{\dagger(i)}$ now has $D$ instead of $\rm{g}$, and the vacuum states are the  
same because the transformation does not mix creation and annihilation operators. We can assume $D_j=0$ for $j\leq j_1$, and the  
remaining eigenvalues to be nonzero. By interchanging particles with holes ($\xi_j^{(0)}\leftrightarrow\xi_j^{\dagger(0)}$) for  
every index such that $D_j=0$ we obtain a linear transformation with an invertible matrix. The calculation of the echo follows the  
same steps as before, except that it is necessary to treat indices $j\leq j_1$ separately. After the Gaussian integration, we are left with  
an expression of the form (\ref{eq:Gaussianinttodet}) in which only indices $j>j_1$ appear. The desired result can be achieved by  
conveniently introducing some rows and columns in the matrix, in such a way to include the parts of the matrices with $j\leq j_1$  
without changing the value of the determinant. 
 
The present derivation cannot be easily extended to more general cases, with two different evolution operators as in (\ref{eq:echodef}), for in that case there are three sets of operators that must be related. The steps taken in the calculation above then  
lead to a result that involves a sum over an exponential number of terms, and so cannot be efficiently evaluated.

\section{Appendix B} 
 
Here we sketch the derivation of the formula (\ref{eq:secondmainformula}) for the decay of the off-diagonal terms $\rho_{01,10}$. 
We start from expression (\ref{eq:echo0110}), which involves two sets of particle operators: one which we note as $\eta^{(1)}$, diagonalizing the effective  
Hamiltonian $H_{10}$, and the set $\eta^{(0)}$ for the unperturbed chain Hamiltonian $H_{00}$. These two sets are connected by a real Bogoliubov transformation.  

The translation operator can then be written as \cite{Note2}: 
\beq 
T = e^{\frac{2\pi i}{N} \sum_{k=1}^N k \eta^{(0)\dagger}_k \eta^{(0)}_k} 
\eeq 
so that the echo is given by: 
\beq 
L_{01,10} (t) = \left|\bra{E_0} e^{\frac{2\pi i}{N} \sum_{k=1}^N k \alpha^\dagger_k \alpha_k} \ket{E_0}\right|^2 
\eeq 
with  
\beq 
\alpha_k=e^{iH_{10}t} \eta^{(0)}_k e^{-iH_{10}t}.
\eeq 
This leads us to an echo of exactly the same form as that in Appendix A, except that the energy matrix $\Lambda$ is replaced by a  
matrix with the eigenvalues of the translation operator, and the set of particles $\eta^{(1)}_k$ is replaced by the $\alpha_k$. The main  
difference here is given by the fact that the Bogoliubov transformation between the $\alpha_k$ and the $\eta^{(0)}_k$ has time-dependent complex  
coefficients: 
\beq 
\eta_j^{(0)}=\sum_k {\rm g'}_{jk}\alpha_k+{\rm h'}_{jk}\alpha_k^\dagger  
\eeq 
with 
\beqa 
{\rm g'}&=&{\rm g}e^{it\Lambda}{\rm g}^t+{\rm h}e^{-it\Lambda}{\rm h}^t,\\ 
{\rm h'}&=&{\rm g}e^{it\Lambda}{\rm h}^t+{\rm h}e^{-it\Lambda}{\rm g}^t. 
\eeqa 
Following steps similar to those in Appendix A, we finally obtain for the echo the expression (\ref{eq:secondmainformula}): 
\beq 
L_{01,10} (t) = \left|\det({\rm g'} \tau^{-d} {\rm g'}^\dagger + {\rm h'} \tau^d {\rm h'}^\dagger)\right| \nonumber 
\eeq 
where the diagonal matrix $\tau$ contains the eigenvalues of the translation  
operator $T$; the structure is the same as in (\ref{eq:mainformula_previousstep}). In this way the decay of the off-diagonal terms can once more be written as the determinant of an $N\times N$ matrix.  
This expression is not as simple as the final one obtained in Appendix A; the reason for this is that the matrices $\rm{g}\pm\rm{h}$ were  
orthogonal, while the matrices $\rm{g'}\pm \rm{h'}$ do not satisfy a similar condition.

 
\end{document}